\documentclass[10pt]{article} 
\usepackage{amssymb}
\usepackage{amsmath}
\usepackage{amsthm}
\usepackage{latexsym} 
\usepackage[dvips]{epsfig}
\usepackage{eufrak}

\theoremstyle{plain}
\newtheorem{proposition}{Proposition} 

\newtheorem{theorem}{Theorem}

\font\SYM=msbm10 
\newcommand{\Real}{\mbox{\SYM R}}

\font\tenscr=rsfs10 scaled1100
\font\sevenscr=rsfs7 
\font\fivescr=rsfs5 
\skewchar\tenscr='177 
\skewchar\sevenscr='177
\skewchar\fivescr='177 
\newfam\scrfam 
\textfont\scrfam=\tenscr
\scriptfont\scrfam=\sevenscr 
\scriptscriptfont\scrfam=\fivescr

\begin{document}


\title{\textbf{A characterisation of Schwarzschildean initial data}}

\author{Juan Antonio Valiente Kroon \thanks{E-mail address:
 {\tt jav@ap.univie.ac.at}} \\
Institut f\"ur Theoretische Physik,\\ Universit\"at Wien,\\
Boltzmanngasse 5, A-1090 Wien,\\ Austria.}

\maketitle

\begin{abstract}
A theorem providing a characterisation of Schwarzschildean initial
data sets on slices with an asymptotically Euclidean end is
proved. This characterisation is based on the proportionality of the
Weyl tensor and its D'Alambertian that holds for some vacuum Petrov
Type D spacetimes (e.g. the Schwarzschild spacetime, the C-metric, but
not the Kerr solution). The 3+1 decomposition of this proportionality
condition renders necessary conditions for an initial data set to be a
Schwarzschildean initial set. These conditions can be written as
quadratic expressions of the electric and magnetic parts of the Weyl
tensor ---and thus, involve only the freely specifiable data. In order
to complete our characterisation, a study of which vacuum static
Petrov type D spacetimes admit asymptotically Euclidean slices is
undertaken. Furthermore, a discussion of the ADM 4-momentum for
boost-rotation symmetric spacetimes is given. Finally, a
generalisation of our characterisation, valid for Schwarzschildean
hyperboloidal initial data sets is put forward.
\end{abstract}

\textbf{Pacs:} 04.20.Ex, 04.20.Jb, 04.20.Ha.

\sloppy

\section{Introduction}
This article is concerned with answering the following question: given
a 3-dimensional manifold, $\mathcal{S}$, and a pair $(h_{ij},K_{ij})$
of symmetric tensors on $\mathcal{S}$ satisfying the Einstein vacuum
constraint equations
\begin{eqnarray}
&& r+ K^2-K_{ij}K^{ij}=0, \\
&& D^jK_{ij}-D_iK=0,
\end{eqnarray}
how do we know that the triplet
$(\mathcal{S},h_{ij},K_{ij})$ corresponds to a slice of the
Schwarzschild spacetime? Above, as well as in the sequel, $D$ and $r$
denote, respectively, the connection and the Ricci scalar of the
3-metric $h_{ij}$, and we have written $K=K^i_{\phantom{i}i}$ for the
trace of extrinsic curvature $K_{ij}$.

The problem stated above is of interest because although the Schwarzschild
spacetime is, arguably, fairly well understood, several aspects of its
3+1 decomposition ---relevant for numerical investigations--- are
still open. Among what is known, one should mention the examples of
time asymmetric slices given by Reinhardt and Estabrook et
al., \cite{Rei73,EstWahlChrDeWSmaTsi73}, and the CMC slicing found by
Beig \& O'Murchadha \cite{BeiOMu98}. Examples of foliations with a
harmonic time function have been given in \cite{SchBauCooShaTeu98},
and conditions for the embedding of spherically symmetric slices in a
Schwarzschild spacetime have been considered in \cite{OmuRos03}.  On the
other hand, however, boosted slices in the Schwarzschild spacetime
constitute , essentially, an uncharted territory. It is not known, for
example, if there are boosted slices which are maximal ---the
available examples, e.g. that given by York in \cite{Yor80}, are
not. That these slices cannot be boosted can be proved by the methods
used in \cite{Val04c}.

We note that in the case of the Minkowski spacetime, the Codazzi
equations readily provide a pointwise ---i.e local--- answer to the
analogue question. Namely, a pair $(h_{ij},K_{ij})$ of symmetric
tensors correspond (locally) to the first and second fundamental form
of a slice $\mathcal{S}$ in Minkowski spacetime if and only if
\begin{subequations}
\begin{eqnarray}  
&& D_{[i}K_{j]l}=0, \\
&& r_{ijkl}=-2 K_{k[i}K_{j]l},
\end{eqnarray}
\end{subequations}
where $D_i$ and $r_{ijkl}$ denote, respectively, the connection and
the Riemann tensor associated to the 3-metric $h_{ij}$.

If the spacetime has a non-vanishing curvature, the
situation is fundamentally more complicated, and in order to obtain a
local answer in a systematic way, one would have to resort to some
---yet unavailable--- 3+1 formulation of the equivalence problem. 

Almost any invariant characterisation of the Schwarzschild spacetime
has to make use, \emph{a fortiori}, of the fact that it is of Petrov
type D ---see e.g. \cite{Kar80} and \cite{FerSae98} \footnote{The
Petrov classification is an algebraic characterisation of the Weyl
tensor based on the solutions of a certain eigenvalue problem. In
particular, a spacetime is said to be of Petrov type D if there are
two vectors $k^\mu$ and $l^\mu$ ---\emph{the principal null
directions}--- such that
\[
C_{\mu\nu\lambda[\rho}k_{\sigma]}k^\nu k^\lambda=0, \quad C_{\mu\nu\lambda[\rho}l_{\sigma]}l^\nu l^\lambda=0.
\]
For further details on the theory of the Petrov classification see
 e.g. \cite{SKMHH}.  }. However, the Petrov type, a neat 4-dimensional
 property of spacetime, tends to project into complicated
 expressions when attempting a 3+1 decomposition of its defining
 relations. The point is, then, to find a description ---if any---of
 the fact that a spacetime is of Petrov type D with a neat 3+1
 decomposition. A description of the desired sort is given by a
 proportionality relation between the D'Alambertian of the Weyl
 tensor and the Weyl tensor itself satisfied by some vacuum Petrov
 type D spacetimes ---Schwarzschild included--- found by Zakharov
 \cite{Zak65,Zak70,Zak72} ---see equation (\ref{zakharov_property}).

\medskip
In what follows, by \emph{the Schwarzschild spacetime} it will be
understood the Schwarzschild-Kruskal maximal extension,
$(\mathcal{M},g_{\mu\nu})$, of the Schwarzschild spacetime
\cite{Kru60}. Accordingly, by a \emph{slice} of the Schwarzschild
spacetime it will be understood that there exists an embedding
$\phi:\mathcal{S}\longrightarrow \mathcal{M}$ such that
$h_{ij}=(\phi^*g)_{ij}$, and
$K_{ij}=\frac{1}{2}(\phi^*\mathcal{L}_nh)_{ij}$, where $n^\mu$ is the
(timelike) $g-$unit normal of $\phi(\mathcal{S})$, $\mathcal{L}$ is
the Lie derivative, and $\phi^*$ denotes the pull back of tensor
fields from $\mathcal{M}$ to $\mathcal{S}$. Furthermore, let
$C_{\mu\nu\lambda\rho}$ denote the Weyl tensor of the metric
$g_{\mu\nu}$, and denote by $E_{\mu\nu}$ and $B_{\mu\nu}$,
respectively, the $n-$electric and $n-$magnetic parts of
$C_{\mu\nu\lambda\rho}$. As $E_{\mu\nu}$ and $B_{\mu\nu}$ are spatial
tensors, we shall be writing $E_{ij}=(\phi^*E)_{ij}$ and
$B_{ij}=(\phi^*B)_{ij}$ ---the tensors $E_{ij}$ and $B_{ij}$ can be
expressed purely in terms of $h_{ij}$ and $K_{ij}$.

In terms of the above language, the answer we want to provide to the
question raised in the opening paragraph is given by the following

\begin{theorem}
\label{main_theorem}
Let $\mathcal{S}$ be a 3-manifold with at least one asymptotically
Euclidean flat end, and let $(h_{ij},K_{ij})$ be a solution to the
Einstein vacuum constraint equations decaying on the asymptotically Euclidean end as
\begin{equation} \label{decay_conditions}
h_{ij}-\delta_{ij}=\mathcal{O}_k(r^{-\beta}), \quad K_{ij}=\mathcal{O}_k(r^{-1-\beta}),
\end{equation}
for some $k\geq 2$ and $\beta>1/2$. Let the ADM 4-momentum
associated to the asymptotic end be non-vanishing. If there is a
function $\alpha$ such that
\begin{subequations}
\begin{eqnarray}
&& 6\left(E_{ik}E^k_{\phantom{k}j}-\frac{1}{3}h_{ij}E^{kl}E_{kl}\right)-6\left(B_{ik}B^k_{\phantom{k}j}-\frac{1}{3}h_{ij}B^{kl}B_{kl}\right)=\alpha E_{ij}, \label{condition_electric}\\ 
&& 12\left(E_{\phantom{k}(i}^kB_{j)k}-\frac{1}{3}h_{ij}E_{kl}B^{kl}\right)=\alpha B_{ij}, \label{condition_magnetic}
\end{eqnarray}
\end{subequations}
then the triplet $(\mathcal{S},h_{ij},K_{ij})$ corresponds to a
(spacelike) slice of the Schwarzschild spacetime. Conversely, for any
slice of the Schwarzschild spacetime the conditions (\ref{condition_electric})
and (\ref{condition_magnetic}) hold with
\begin{equation}
\alpha=-\frac{6m}{r^3},
\end{equation} 
where $r$ is the radial coordinate in the standard Schwarzschild coordinates.
\end{theorem}

In the previous theorem by an \emph{asymptotically Euclidean end} it
is understood a portion of $\mathcal{S}$ which is diffeomorphic to
\begin{equation}
\left\{ x^i\in \Real^3 \;\bigg|\;|x|=\left(\sum^3_{i=1} (x^i)^2\right)^{1/2}>r_0\right\},
\end{equation}
where $r_0$ is some positive real number. Note that the (spacelike)
slices covered by the latter theorem are not necessarily Cauchy
hypersurfaces. However, hyperboloidal hypersurfaces not intersecting
one of the two spatial infinities of the Kruskal extension are
excluded ---see figure 1. 

The decay conditions (\ref{decay_conditions}) with the prescribed
values of the constants $k$ and $\beta$ are of technical nature. The
notation $\mathcal{O}_k$ is explained in appendix A. Among other
things, they ensure that ---see e.g. \cite{Bar86,Chr86}--- the ADM
4-momentum \cite{ArnDesMis62} given via the integrals
\begin{subequations}
\begin{eqnarray}
&& p_0=\frac{1}{16\pi} \int_{S_\infty} \left (\partial_j h_{ij}-\partial_i h \right) dS^i, \\
&& p_i=\frac{1}{8\pi}\int_{S_\infty}\left(K_{ij} -K\delta_{ij}\right)dS^j,
\end{eqnarray}
\end{subequations}  
where $h=h_{ij}\delta^{ij}$  is well defined.

\begin{figure}[t]
\centering
\includegraphics[width=.7\textwidth]{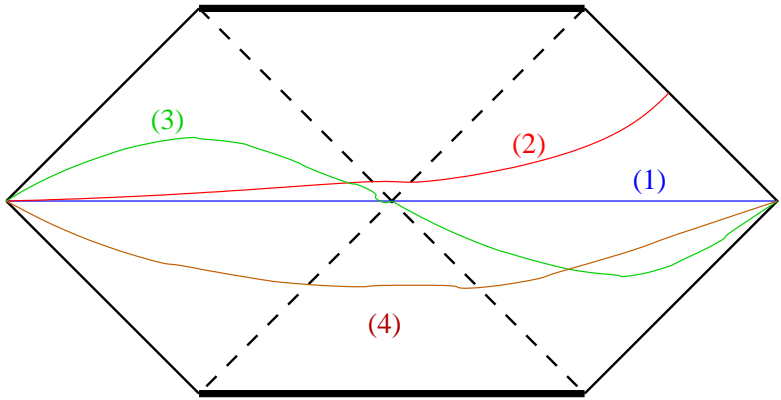}
\caption{Schematic description of types of hypersurfaces covered by
our main theorem: (1) a time symmetric Cauchy hypersurface, (2) a
hyperboloid which also reaches one of the two spatial infinities, (3)
a boosted slice, (4) a generic non-time symmetric Cauchy
hypersurface.}
\end{figure}

Given a hypersurface $\mathcal{S}$ satisfying the conditions
(\ref{condition_electric}) and (\ref{condition_magnetic}), the
assumption of the existence of an asymptotically flat end with a
non-vanishing ADM mass is sharp in order to be able to single out
Schwarzschildean data. If, for example, no statement is made about the
ADM momentum, then the initial data set can be either a
Schwarzschildean one, or one corresponding to the C-metric. In this
sense, our characterisation contains a global element. In order to
obtain a purely local characterisation of Schwarzschild data, one
would have to undertake, for example, a 3+1 decomposition of the
characterisation of the Schwarzschild spacetime in terms of
concomitants of the Weyl tensor obtained by Ferrando \& S\'{a}ez
\cite{FerSae98} ---this will be presented elsewhere.

\bigskip
The article is structured as follows: in section 2, we discuss the
property of the D'Alambertian of the Weyl tensor of some vacuum Petrov
type D spacetimes which is the keystone of our characterisation ---the
\emph{Zakharov property}. A relation of the Petrov type D spacetimes
satisfying this property is given. In section 3 we consider the 3+1
decomposition of the Zakharov property. In section 4 a discussion of
which vacuum static Petrov type D spacetimes admit  asymptotically
Euclidean slices is given. Section 5 is concerned with the ADM
4-momentum of boost-rotation symmetric spacetimes. Finally, in section
6 the main results of the previous sections ---propositions
\ref{the_ones}, \ref{Z_projections}, \ref{static_aflat},
\ref{ADM_mass}--- are recalled and put in context to render our main
result, theorem \ref{main_theorem}. The shortcomings of our
characterisation are discussed briefly, and a generalisation of the
characterisation, valid for hyperboloidal data is given ---see theorem
\ref{hyperboloid_theorem}. There is, also, an appendix is which some
notation issues are addressed.

\section{A result on type D spacetimes}

Let $R_{\mu\nu\lambda\rho}$ denote the Riemann tensor of the metric
$g_{\mu\nu}$. Our point of departure is the following curious result
to be found in the Exact Solutions book \cite{SKMHH}: 

\begin{theorem}[Zakharov 1965, 1970, 1972] \label{theorem:zakharov}
Vacuum fields satisfying the equation
\begin{equation}
\label{zakharov_property}
R_{\mu\nu\lambda\rho;\sigma}^{\phantom{\mu\nu\lambda\rho;\sigma}\sigma}=\alpha R_{\mu\nu\lambda\rho}, 
\end{equation}
for a certain function $\alpha$ are either type N ($\alpha=0$) or type
D ($\alpha \neq 0$).
\end{theorem}

The proof of this theorem follows immediately from $R_{\mu\nu}=0$ and
the identity \cite{Zak72}
\begin{equation} \label{the_identity}
R_{\mu\nu\lambda\rho;\sigma}^{\phantom{\mu\nu\lambda\rho;\sigma}\sigma}=R^\sigma_{\phantom{\sigma}\tau \mu\nu}R^\tau_{\phantom{\tau}\sigma \lambda\rho}+2(R^\sigma_{\phantom{\sigma}\mu\rho \tau}R^\tau_{\phantom{\tau}\lambda\nu \sigma}-R^\sigma_{\phantom{\sigma}\nu\rho \tau}R^\tau_{\phantom{\tau}\lambda\mu \sigma}),
\end{equation}
written down with respect to a principal tetrad ---see
\cite{Zak65,Zak70}. The theorem \ref{theorem:zakharov} stems from
attempts due to A.L. Zel'manov ---in the case $\alpha=0$--- of obtaining a
characterisation of spacetimes containing gravitational radiation.

\medskip
A direct evaluation shows that the property (\ref{zakharov_property})
---which we shall call the \emph{Zakharov property}--- is satisfied by
the Schwarzschild spacetime, but for example, not by the Kerr
solution. As the vacuum Petrov type D spacetimes are all known thanks
to the work of Kinnersley \cite{Kin69}, it is not too taxing to
perform a casuistic analysis to see which are the ones satisfying the
property (\ref{zakharov_property}). Kinnersley's analysis made use of
the Newman-Penrose (NP) formalism \cite{NewPen62} and divides
naturally into two cases: those solutions for which the NP spin
coefficient $\rho$ ---the expansion--- vanishes and those for which it
does not. The case with $\rho\neq 0$ divides, in turn, into 9
subcases. The solutions in case I have, in general, a non-vanishing
NUT parameter, $l$. If $l=0$, then one obtains the Ehlers-Kundt
solutions A1, A2 and A3 ---see \cite{EhlKun62}. These solutions are
static, and save the solution A1 (Schwarzschild) they are not
asymptotically flat in the sense that there are no constants $k \geq
2$, $\beta>1/2$ for which
\begin{equation} \label{non_aflat}
g_{\mu\nu}-\eta_{\mu\nu}=\mathcal{O}_k(r^{-\beta}).
\end{equation}
The latter definition of asymptotic flatness has been borrowed from
\cite{BeiChr96,BeiChr97a} and will turn out to be most convenient for
our endeavours.  The $\rho\neq 0$ case II.A to II.F contain the
Kerr-NUT solution and also other (non-asymptotically flat)
solutions describing spinning bodies. The cases $\rho\neq 0$ III.A and
III.B correspond, respectively, to the C-metric and its
generalisation, the spinning C-metric. These solutions are known to be
compatible (for particular ranges of the parameters) with the notion
of asymptotic flatness ---see
\cite{AshDra81,BicPra99b,PraPra98}. Finally, the solutions with
$\rho=0$ divide, in turn, in two classes A and B. The class A
corresponds to the Ehlers-Kundt solutions B1 to B3 and are not
asymptotically flat in the sense given by equation
(\ref{non_aflat}). The solutions of class A are spinning
generalisations of class B. A summary of which of the vacuum 
Petrov type D spacetimes satisfy the Zakharov property, equation
(\ref{zakharov_property}), is given in table 1. From there, we derive
the following

\begin{proposition}
\label{the_ones}
The only type D solutions satisfying the Zakharov property, equation
(\ref{zakharov_property}), are those with hypersurface orthogonal
Killing vectors ---that is, the Ehlers-Kundt solutions A1, A2, A3, B1,
B2, B3 and C.
\end{proposition}

\begin{table}[t]
\center
\begin{tabular}{|l|l|l|}\hline
$\rho \neq 0$ &  case I (NUT metrics including Schwarzschild) & only if $l=0$\\ \cline{2-3}
$\phantom{\rho \neq 0}$ & case II.A (Kerr-NUT)& no \\
$\phantom{\rho \neq 0}$ & case II.B & no    \\ 
$\phantom{\rho \neq 0}$ & case II.C & no   \\
$\phantom{\rho \neq 0}$ & case II.D & no    \\
$\phantom{\rho \neq 0}$ & case II.E & no   \\
$\phantom{\rho \neq 0}$ & case II.F & no   \\ \cline{2-3}
$\phantom{\rho \neq 0}$ & case III.A (C-metric) & yes \\
$\phantom{\rho \neq 0}$ & case III.B (twisting C-metric) & no \\ \hline
$\rho =0$&  case A & yes \\
$\phantom{\rho =0}$&  case B & no\\ \hline
\end{tabular}
\caption{Relation of the vacuum, type D spacetimes satisfying the
Zakharov property. The description of the different cases follows the
discussion given in Kinnersley's analysis ---see \cite{Kin69} and also
\cite{NewTamUnt63}. The case I with $l=0$ corresponds to Ehlers-Kundt
solutions A1 (Schwarzschild), A2 and A3. The case A corresponds to the
Ehlers-Kundt solutions B1, B2 and B3 \cite{EhlKun62}.}
\end{table}

\bigskip
Arguably, of the spacetimes in table 1 satisfying the property
(\ref{zakharov_property}) those of most interest are the
Schwarzschild spacetime and the C-metric. For the Schwarzschild
spacetime in the standard coordinates $(t,r,\theta,\varphi)$ 
the line element assumes the form
\begin{equation}
g_S=\left(1-\frac{2m}{r}\right)dt^2-\left(1-\frac{2m}{r}\right)^{-1}dr^2-r^2(d\theta^2 +\sin^2\theta d\varphi^2),
\end{equation}
nd the proportionality function is given by
\begin{equation}
\alpha_S=-\frac{6m}{r^3}.
\end{equation}
On the other hand, for the C-metric in the coordinates $(t,x,y,p)$
---see e.g. \cite{KinWal70}--- such that
\begin{equation}
g_C=\frac{1}{A^2(x+y)^2}\left(F(y)dt^2 -\frac{dx^2}{G(x)}-\frac{dy^2}{F(y)}-G(x)dp^2\right),
\end{equation}
where 
\begin{equation}
G(x)=1-x^2-2mAx^3, \qquad F(y)=-1+y^2-2mAy^3,
\end{equation}
one has that
\begin{equation}
\alpha_C=-6A^3m(x+y)^3.
\end{equation}

\section{A $3+1$ decomposition}
The property (\ref{zakharov_property}) in theorem
\ref{theorem:zakharov} provides the cornerstone for a characterisation of
the Schwarzschild spacetime that projects neatly under a 3+1
decomposition. The crucial observation is that in vacuum, the tensor
\begin{equation}
Z_{\mu\nu\lambda\rho}=R_{\mu\nu\lambda\rho;\sigma}^{\phantom{\mu\nu\lambda\rho;\sigma}\sigma}=C_{\mu\nu\lambda\rho;\sigma}^{\phantom{\mu\nu\lambda\rho;\sigma}\sigma},
\end{equation}
where $C_{\mu\nu\lambda\rho}$ is the Weyl tensor of $g_{\mu\nu}$, is
Weyl-like ---that is, it is tracefree;
$Z_{\mu\nu\lambda\rho}=Z_{\lambda\rho\mu\nu}=-Z_{\nu\mu\lambda\rho}=-Z_{\mu\nu\rho\lambda}$;
and satisfies the first Bianchi identity
$Z_{\mu\nu\lambda\rho}+Z_{\lambda\mu\nu\rho}+Z_{\nu\lambda\mu\rho}=0$.

Let $n^\mu$ be an unit timelike vector, and let us denote by
$h^\mu_{\phantom{\mu}\nu}= g^\mu_{\phantom{\mu}\nu}-n^\mu
n_\nu=\delta^\mu_{\phantom{\mu}\nu}-n^\mu n_\nu$ the associated
projector. Following the notation and conventions of \cite{Fri96}, we
decompose the Weyl tensor as
\begin{equation}
C_{\mu\nu\lambda\rho}=2\bigg(l_{\mu[\lambda}E_{\rho]\nu}-l_{\nu[\lambda}E_{\rho]\mu}-n_{[\lambda}B_{\rho]\tau}\epsilon^\tau_{\phantom{\tau}\mu\nu}-n_{[\mu}B_{\nu]\tau}\epsilon^\tau_{\phantom{\tau}\lambda\rho} \bigg),
\end{equation}
where
\begin{equation}
E_{\tau\sigma}=C_{\mu\nu\lambda\rho}h^\mu_{\phantom{\mu}\tau}n^\nu h^\lambda_{\phantom{\lambda}\sigma} n^\rho, \quad B_{\tau\sigma}=C^*_{\mu\nu\lambda\rho}h^\mu_{\phantom{\mu}\tau} n^\nu h^\lambda_{\phantom{\lambda}\sigma}n^\rho,
\end{equation}
denote, respectively, the \emph{n-electric} and \emph{n-magnetic}
parts of $C_{\mu\nu\lambda\rho}$,
$\epsilon_{\tau\lambda\rho}=\epsilon_{\sigma\tau^\prime\lambda^\prime\rho^\prime}n^\sigma
h^{\tau^\prime}_{\phantom{\tau^\prime}\tau}
h^{\lambda^\prime}_{\phantom{\lambda^\prime}\lambda}
h^{\rho^\prime}_{\phantom{\rho^\prime}\rho}$ is the spatial
Levi-Civita tensor, $l_{\mu\nu}=h_{\mu\nu}+n_\mu n_\nu$, and 
$C^*_{\mu\nu\lambda\rho}=\frac{1}{2}C_{\mu\nu\tau\sigma}\epsilon^{\tau\sigma}_{\phantom{\tau\sigma}\lambda\rho}$
denotes the dual of $C_{\mu\nu\lambda\rho}$. The electric and magnetic
parts of $C_{\mu\nu\lambda\rho}$ are symmetric,
$E_{\mu\nu}=E_{\nu\mu}$, $B_{\mu\nu}=B_{\nu\mu}$, and traceless
$E^\mu_{\phantom{\mu}\mu}=B^\mu_{\phantom{\mu}\mu}=0$. Moreover, they
are spatial tensors in the sense that $E_{\mu\nu^\prime}
h^{\nu^\prime}_{\phantom{\nu^\prime}\nu}=B_{\mu\nu^\prime}
h^{\nu^\prime}_{\phantom{\nu^\prime}\nu}=0$; and
$C_{\mu\nu\lambda\rho}=0$ if and only if $E_{\mu\nu}=B_{\mu\nu}=0$.

Using the embedding $\phi$, we can calculate the pull-backs of the
electric and magnetic parts of $C_{\mu\nu\lambda\rho}$ to the
hypersurface $\mathcal{S}$. Consequently, let us write
$E_{ij}=(\phi^*E)_{ij}$ and $B_{ij}=(\phi^*B)_{ij}$. It is a direct
consequence of the Codazzi equations that one can write
\begin{subequations}
\begin{eqnarray}
&& E_{ij}=r_{ij}+K K_{ij}-K_{ik}K^k_{\phantom{k}j}, \\
&& B_{ij}=-2\epsilon_{i}^{\phantom{i}kl}D_{k}K_{lj},
\end{eqnarray}
\end{subequations}
where $r_{ij}$ denotes the Ricci tensor of the 3-metric
$h_{ij}=(\phi^*h)_{ij}$. Thus, on $\mathcal{S}$, the electric and
magnetic parts of the Weyl tensor can be entirely written in terms of
the initial data $(h_{ij},K_{ij})$. Note, that in particular, for time
symmetric spacetimes one has $B_{ij}=0$ as $K_{ij}=0$.

The tensor $Z_{\mu\nu\lambda\rho}$, being Weyl-like, admits a similar
decomposition in terms of $n$-electric and $n$-magnetic parts, which
we shall denote by $D_{\mu\nu}$ and $H_{\mu\nu}$, respectively. Hence, we write
\begin{equation}
Z_{\mu\nu\lambda\rho}=2\bigg(l_{\mu[\lambda}D_{\rho]\nu}-l_{\nu[\lambda}D_{\rho]\mu}-n_{[\lambda}H_{\rho]\tau}\epsilon^\tau_{\phantom{\tau}\mu\nu}-n_{[\mu}H_{\nu]\tau}\epsilon^\tau_{\phantom{\tau}\lambda\rho} \bigg),
\end{equation}
where
\begin{equation}
D_{\tau\sigma}=Z_{\mu\nu\lambda\rho}h^\mu_{\phantom{\mu}\tau}n^\nu h^\lambda_{\phantom{\lambda}\sigma} n^\rho, \quad H_{\tau\sigma}=Z^*_{\mu\nu\lambda\rho}h^\mu_{\phantom{\mu}\tau} n^\nu h^\lambda_{\phantom{\lambda}\sigma}n^\rho,
\end{equation}
and $Z^*_{\mu\nu\lambda\rho}=\frac{1}{2}Z_{\mu\nu\tau\sigma}\epsilon^{\tau\sigma}_{\phantom{\tau\sigma}\lambda\rho}$. As in the case of $E_{\mu\nu}$ and $B_{\mu\nu}$, one has that $D_{\mu\nu}=D_{\nu\mu}$, $H_{\mu\nu}=H_{\nu\mu}$, $D^\mu_{\phantom{\mu}\mu}=H^\mu_{\phantom{\mu}\mu}=0$,  $D_{\mu\nu^\prime}
h^{\nu^\prime}_{\phantom{\nu^\prime}\nu}=H_{\mu\nu^\prime}
h^{\nu^\prime}_{\phantom{\nu^\prime}\nu}=0$; and
$Z_{\mu\nu\lambda\rho}=0$ if and only if $D_{\mu\nu}=H_{\mu\nu}=0$.

For vacuum spacetimes, the identity (\ref{the_identity}) allows to
write the tensors $D_{\mu\nu}$ and $H_{\mu\nu}$ as quadratic
expressions of $E_{\mu\nu}$ and $B_{\mu\nu}$. A lengthy, but
straightforward calculation renders the remarkably simple expressions:
\begin{subequations}
\begin{eqnarray}
  &&D_{\mu\nu}=6\left(E_{\mu \sigma}E^\sigma_{\phantom{\sigma}\nu}-\frac{1}{3}h_{\mu\nu}E^{\sigma\tau}E_{\sigma\tau}\right)-6\left(B_{\mu\sigma}B^\sigma_{\phantom{\sigma}\nu}-\frac{1}{3}h_{\mu\nu}B^{\sigma\tau}B_{\sigma\tau}\right), \\
  &&H_{\mu\nu}=12\left(E_{\phantom{\sigma}(\mu}^\sigma B_{\nu)\sigma}-\frac{1}{3}h_{\mu\nu}E_{\sigma\tau}B^{\sigma\tau}\right).
\end{eqnarray}
\end{subequations}
These expressions can be pulled-back to the hypersurface
$\mathcal{S}$ by means of the embedding $\phi$ to obtain the
following

\begin{proposition}\label{Z_projections}
Necessary conditions for an initial data set
$(\mathcal{S},h_{ij},K_{ij})$ to be a Schwarzschildean initial data
set are:
\begin{subequations}
\begin{eqnarray}
&&  D_{ij}=\alpha E_{ij}, \\ 
&&  H_{ij}=\alpha B_{ij},
\end{eqnarray}
\end{subequations}
where $\alpha=-6m/r^3$, where $r$ is the standard Schwarzschild radial
coordinate.
\end{proposition} 

Note that the C-metric satisfies an analogous theorem with
$\alpha=-6Am(x+y)^3$.

\section{Asymptotic flatness and static type D spacetimes}

In order to be able to discern Schwarzschildean data from among all
those vacuum type D initial data sets satisfying the conditions
$D_{ij}=\alpha E_{ij}$ and $H_{ij}=\alpha B_{ij}$, we require a couple
of further results. Our first task is to get rid of those spacetimes
which admit no slices with asymptotically Euclidean ends. Intuitively,
it seems clear that a static spacetime which is not asymptotically
flat should not admit slices with asymptotically flat ends. More
precisely, one has the following

\begin{proposition} \label{static_aflat}
If a vacuum static spacetime is not asymptotically flat in the sense
given by equation (\ref{non_aflat}) ---i.e. it belongs to the
Ehlers-Kundt classes A2, A3 or B1, B2, B3--- then it admits no slices
with asymptotically Euclidean ends for which the decay conditions
(\ref{decay_conditions}) hold.
\end{proposition}

It can be readily checked by direct computation that the spacetimes of
the Ehlers-Kundt classes A2, A3 or B1, B2, B3 are not asymptotically
flat in the sense discussed in the introduction. The proof of the
proposition is by contradiction. Assume that our non-asymptotically
flat, static spacetime, $\mathcal{M}$, admits a slice, $\mathcal{S}$,
with an asymptotically Euclidean end for which the asymptotic decay
conditions (\ref{decay_conditions}) hold. By construction, in this
slice one has that $h_{ij}-\delta_{ij}=\mathcal{O}_k(r^{-\beta})$ and
$K_{ij}=\mathcal{O}_{k-1}(r^{-1-\beta})$ with $\beta >1/2$ and $k\geq
2$. For this type of initial data the solution to the \emph{boost
problem} ---see \cite{ChrOMu81}--- ensures the existence of a
\emph{boost-type domain} $\Omega_{r_0,\theta}$ of the form
\begin{equation} \label{boost_domain}
\Omega_{r_0,\theta}=\big\{ (t,x^i)\in \Real \times \Real^3 \; \big | \; |x|\geq r_0, \; |t|\leq \theta |x| \big \},
\end{equation}
for some constants $r_0$ and $\theta$, such that $r_0>0$ and
$0<\theta<1$. From the fact that $\mathcal{M}$ is static, it follows
that the slice $\mathcal{S}$ possesses a static \emph{Killing initial data set}
(KID). That is, there exists a pair $(N,X^\mu)$, where $N$ is a scalar
field and $X^\mu$ is a spatial vector field ($X^\mu h_{\mu\nu}=0$)
such that $\xi^\mu|_\mathcal{S}=Nn^\mu+X^\mu$, where $\xi^\mu$ denotes
the static Killing vector of the spacetime $\mathcal{M}$, and $n^\mu$
is the normal to $\mathcal{S}$. In what follows, let $X^i$ denote the
pull-back of $X^\mu$, i.e. $X^i=(\phi^*X)^i$. Now, it is natural to
consider the evolution of the initial data set $(h_{ij},K_{ij})$ along
the flow given by the static Killing vector $\xi^\mu$. Thus, in
$\Omega_{r_0,\theta}$ one has that the spacetime metric is given by
\begin{equation}
g= -N^2 dt^2 + h_{ij}(dx^i+X^i)(dx^j+X^j).
\end{equation}
Recall that along this flow one has that $\partial_t
h_{ij}=0$. Furthermore ---see theorem 2.1 in \cite{BeiChr97a} and also
theorem 2.1 in \cite{BeiChr96}--- the lapse $N$ and shift $X^i$ behave
asymptotically as
\begin{subequations}
\begin{eqnarray}
&& N=1+\mathcal{O}_k(r^{-\beta}), \\
&& X^i=\mathcal{O}_k(r^{-\beta}),
\end{eqnarray}
\end{subequations}
with $\beta > 1/2$ and $k\geq 2$. Thus, it follows that in
$\Omega_{r_0,\theta}$
\begin{equation}
g_{\mu\nu}-\eta_{\mu\nu}=\mathcal{O}_k(r^{-\beta}).
\end{equation}
This is a contradiction to the assumption
that spacetime is not asymptotically flat.

\section{The ADM mass of the C-metric}

The proposition \ref{static_aflat} reduces our task of characterising
Schwarzschildean initial data to finding a way of distinguishing 
between initial data corresponding to the C-metric and those
corresponding to the Schwarzschild spacetime. 

The C-metric belongs to the so-called boost-rotation symmetric
spacetimes ---see \cite{Bon83,BicSch89a,PraPra98}---, that is, it
possesses two commuting, hypersurface orthogonal Killing vectors. One
of them is axial, and the other is of boost type. An argument outlined by
Dray in \cite{Dra82} leads to

\begin{proposition}[Dray, 1982]
\label{ADM_mass}
The ADM 4-momentum of a boost-rotation symmetric spacetimes which is
asymptotically flat ---in the sense of equation (\ref{non_aflat})---
vanishes.
\end{proposition}

Our strategy will be to make use of the latter result to discern
between initial data sets corresponding to the C-metric, and those of
the Schwarzschild spacetime.

Dray's original argument lacks of some technical details, which we now
proceed to fill. Let $(\mathcal{M},g)$ denote a boost-rotation
symmetric spacetime, and let us denote by $\chi^\mu$, $\xi^\mu$,
respectively the axial and boost Killing vectors of the spacetime. The
vectors $\chi^\mu$ and $\xi^\mu$ commute. From the general theory of
boost-rotation symmetric spacetimes given in \cite{BicSch89a} we know
that there is a region of the spacetime ---the one below the so-called
\emph{roof}--- where the spacetime is static. The portion of the
spacetime below the roof admits a boost-type domain,
$\Omega_{r_0,\theta}$, like the one in (\ref{boost_domain}). From the
fact that static spacetimes ---and, in general, flat stationary
spacetimes--- admit a smooth null infinity ---see
e.g. \cite{Dai01b}--- and from the analysis of \cite{BeiChr97a} it
follows that on $\Omega_{r_0,\theta}$ there exist matrices
$\sigma_{\mu\nu}=\sigma_{[\mu\nu]}$, $\rho_{\mu\nu}=\rho_{[\mu\nu]}$
such that
\begin{subequations}
\begin{eqnarray}
&& \chi^\mu-\sigma^\mu_{\phantom{\mu}\nu}x^\nu=\mathcal{O}_k(r^{-\beta}),\label{asymptotic_form_Killings_a} \\
&& \xi^\mu-\rho^\mu_{\phantom{\mu}\nu}x^\nu=\mathcal{O}_k(r^{-\beta}), \label{asymptotic_form_Killings_b} 
\end{eqnarray}
\end{subequations}
for $k\geq 2$ and $\alpha>1/2$,  with $\sigma^\mu_{\phantom{\mu}\nu}\equiv
\eta^{\mu\lambda}\sigma_{\lambda\nu}$,
$\rho^\mu_{\phantom{\mu}\nu}\equiv
\eta^{\mu\lambda}\rho_{\lambda\nu}$, and $\eta_{\mu\nu}$ denoting the
Minkowski metric. Without loss of generality assume that the axis of symmetry
of the axial Killing vector lies along the $x^3$ axis. Accordingly,
\begin{subequations}
\begin{eqnarray}
&& \sigma_{\mu\nu}x^\nu=(0,-x^2,x^1,0), \\
&&  \rho_{\mu\nu}x^\nu=(-x^3,0,0,-t).
\end{eqnarray}
\end{subequations}
 Thus, from the
commuting nature of the two Killing vectors $\chi^\mu$ and $\xi^\mu$
it follows that
\begin{equation}
\sigma^\mu_{\phantom{\mu}\nu}=\begin{pmatrix} 0 & 0 & 0 & 0 \\
                                       0 & 0 & -1 &  0 \\
                                       0 & 1 & 0 &  0 \\
                                       0 & 0 & 0 &  0
                        \end{pmatrix}, \quad
\rho^\mu_{\phantom{\mu}\nu}=\begin{pmatrix} 0 & 0 & 0 & -1 \\
                                       0 & 0 & 0 &  0 \\
                                       0 & 0 & 0 &  0 \\
                                      -1 & 0 & 0 &  0
                        \end{pmatrix}.
\end{equation}
One can associate to the boost-type domain $\Omega_{r_0,\theta}$,
provided that $k \geq 2$ and $\beta >1/2$, in an unique way an ADM
4-momentum vector $p^\mu$ ---see e.g. \cite{Bar86,Chr86}. It follows
from the theory developed in \cite{BeiChr97a} that
\begin{equation}
\sigma^\mu_{\phantom{\mu}\nu}p^\nu=\rho^\mu_{\phantom{\mu}\nu}p^\nu=0,
\end{equation}
whence necessarily
\begin{equation}
p^\mu=0,
\end{equation}
which is the observation made by Dray in \cite{Dra82}. As a side
remark, note that the above result needs not to hold if the Killing
vectors are non-commuting.

\section{Concluding remarks}
Our main theorem ---see the introductory section--- follows directly
from the propositions \ref{the_ones}, \ref{Z_projections},
\ref{static_aflat} and \ref{ADM_mass}. 

It is clear from the argumentation that the conditions to single out
the Schwarzschild solution are sharp. In particular, as seen from
proposition \ref{ADM_mass} if no remark on the ADM 4-momentum is made,
initial data for the C-metric is included. Precisely because of this
condition, it is that our argumentation can not be extended to include
hyperboloidal initial data sets not intersecting spatial infinity like
the ones discussed in \cite{Sch02}. Intuitively, in the case of
hyperboloidal data one would try to replace the condition on the ADM
4-momentum by some condition regarding the Bondi 4-momentum. However,
it is well known that the Bondi mass of boost-rotation symmetric
spacetimes is non-vanishing ---see e.g. \cite{PraPra98}. An
alternative is to replace the condition on the ADM 4-momentum by a
condition on the so-called Newman-Penrose (NP) constants
\cite{NewPen65,NewPen68}. The NP constants vanish
for the Schwarzschild spacetime ---see e.g. \cite{DaiVal02}---, but
are non-vanishing for the C-metric
---cfr. e.g. \cite{LazVal00}. Friedrich \& K\'ann\'ar \cite{FriKan00}
have shown how these quantities defined at null infinity can be
expressed in terms of Cauchy initial data. In principle, the NP
constants are also expressible in terms of hyperboloidal data ---the
details of this have not yet been worked out, and will be pursued
elsewhere. Accordingly, we state ---without going fully into the details--- the following

\begin{theorem}
\label{hyperboloid_theorem}
Let $\mathcal{S}$ be a 3-manifold with a hyperboloidal end, and let
$(h_{ij},K_{ij})$ be a pair of symmetric tensors on $\mathcal{S}$
satisfying the Einstein vacuum constraints. If there is a function
$\alpha$ such that the conditions (\ref{condition_electric}) and
(\ref{condition_magnetic}) hold, i.e.
\begin{equation}
D_{ij}=\alpha E_{ij}, \quad H_{ij}=\alpha B_{ij},
\end{equation}
 then the triplet $(\mathcal{S},h_{ij},K_{ij})$ corresponds to a
 slice of the Schwarzschild spacetime.
\end{theorem}

For a discussion on the appropriate boundary conditions giving rise to
a hyperboloidal end, the reader is remitted to \cite{Fri83} ---see
also \cite{AndChr93} and reference therein.

The question whether the theorems 1 or 2 can be used to construct
Schwarzschildean initial data sets with especial properties ---for
example boosted slices with vanishing mean curvature, if these
exist--- remains open. In any case, the conditions
(\ref{condition_electric}) and (\ref{condition_magnetic}) are
necessary conditions for an initial data set to be
Schwarzschildean. Also, it would be of interest to see if it is
possible to obtain a reformulation of (\ref{condition_electric}) and
(\ref{condition_magnetic}) which does not contain the function
$\alpha$. These ideas will be pursued elsewhere.

\section*{Acknowledgements}
My gratitude is due to R. Beig for stimulating conversations which
lead to the conception of this problem, and for encouragement and
interest while the problem was being worked out. I am also grateful to
CM Losert for a careful reading of the manuscript. This research has
been partly supported by a Lise Meitner fellowship (M814-N02) of the
Fonds zur F\"orderung der wissenschaftliche Forschung (FWF), Austria.

\appendix

\section{The notation $\mathcal{O}_k$}

In this article we follow the $\mathcal{O}_k$ notation introduced in
\cite{BeiChr97a}. Given a function $\phi$ on the boost-type domain
$\Omega_{r_0,\theta}$, we say that $\phi=\mathcal{O}_k(r^\beta)$, for
$\beta\in\Real$, if $\phi\in C^k(\Omega_{r_0,\theta})$ and there is a
function $C(t)$ such that
\[
|\partial_{\alpha_1}\cdots \partial_{\alpha_i}\phi| \leq C(t)(1+|x|)^{\beta-i}, \quad 0\leq i \leq k.
\]


\end{document}